\documentclass[]{pasj01}  
\draft
\usepackage{mathrsfs}
\usepackage{graphicx}
\usepackage{mediabb}
\usepackage{color} 
\usepackage{bm}   
\usepackage[top=3cm, bottom=3cm, left=3cm, right=3cm]{geometry}

\begin{document}
\title{Rotation Curve of the Milky Way and the Dark Matter Density}
 
\author{Yoshiaki Sofue  \\
Institute of Astronomy, The University of Tokyo, Mitaka, Tokyo 181-0015, Japan\\ E-mail: sofue@ioa.s.u-tokyo.ac.jp}

\def\be{\begin{equation}} \def\ee{\end{equation}} 
\def\bc{\begin{center}} \def\ec{\end{center}}  \def\ef{\end{figure}}   
\def\Msun{M_\odot} \def\msun{M_\odot}
\def\msqpc{\Msun {\rm pc}^{-2}} \def\mcupc{\Msun {\rm pc}^{-3}}  
\def\Vsun{ V_0 } \def\Rsun{ R_0 } \def\rzero{R_0} \def\vzero{ V_0 }   
 \def\dv{ de Vaucouleurs }  
\def\vr{v_{\rm r}} \def\vp{v_{\rm p}} \def\vrot{ $V_{\rm rot}$ }  
\def\Vrot{ V_{\rm rot} }  \def\Deg{^\circ} \def\deg{^\circ} 
 \def\lv{(l, v)}   \def\cos{~{\rm cos}~} \def\sin{~{\rm sin}~} 
\def\sinl{~{\rm sin~}l~} \def\cosl{~{\rm cos~}l~}  \def\masy{ mas y$^{-1}$ }
\def\vp{v_{\rm p}} \def\vr{v_{\rm r}} \def\Up{U_{\rm p}} 
\def\Ur{U_{\rm r}} \def\V0{ V_0 }
\def\Mb{M_{\rm b}} \def\Md{M_{\rm d}}  
\def\ab{a_{\rm b}} \def\ad{a_{\rm d}}   
 \def\kms{ km s$^{-1}$ }     
 \def\gevcc{ GeV cm$^{-3}$ } \def\Gevcc{ GeV cm$^{-3}$ }
\def\RE{ R_{\rm E} } \def\rhoE{ \rho_{\rm E} } \def\ME{M_{\rm E} } 
  \def\Gck{\Gevcc (\kms)$^{-1}$}   
\def\log{{\rm log}}
\def\apj{ApJ} \def\mnras{MNRAS} \def\pasj{PASJ} \def\pasp{PASP} \def\apjl{ApJ.L.} \def\aj{AJ} 
\def\dm{\df} \def\prd{Phys. Rev. D} \def\prl{Phys. Rev. Let.} \def\aap{AA} \def\aapr{AA Rev.}  
\def\revise{} 
\def\rev{}
\def\revD{}
      
\maketitle

\begin{abstract}
We review the~current status of the~study of rotation curve (RC) of the~Milky Way, and~present a~unified RC from the~Galactic Center to the galacto-centric distance of about 100 kpc. The~RC is used to directly calculate the~distribution of the~surface mass density (SMD). We then propose a~method to derive the~distribution of dark matter (DM) density in the~in the~Milky Way using the~SMD distribution. The~best-fit dark halo profile yielded a local DM density of \mbox{$\rho_\odot = 0.36\pm 0.02$ \gevcc.} We also review the~estimations of the~local DM density in the~last decade, and~show that the~value is converging to a~value at $\rho_\odot=0.39\pm 0.09$ \gevcc. \\


({\it Invited review accepted for Galaxies to appear in special issue on "Debate on the Physics of Galactic Rotation and the Existence of Dark Matter"})

\end{abstract} 

\section{Introduction}
 
The rotation curve (RC) of the~Milky Way was obtained by observations of galactic objects in the~non-MOND (MOdified Newtonian Dynamics)
frame work. The~existence of the~dark halo (DH) has been confirmed by the~analysis of the observed RCs, assuming that ~Newtonian dynamics applies evenly to the~result of the~observations. In this article, current works of RC observations are briefly reviewed, and~a~new estimation of the~local dark matter (DM) density is presented in the~framework of Newtonian dynamics.  

An RC is defined as the~mean circular velocity $\Vrot$ around the~nucleus plotted as a~function of the~galacto-centric radius $R$. Non-circular streaming motion due to the triaxial mass distribution in a bar is crucial for kinematics in the~innermost region, though it does not affect the~mass determination much in the~disk and~halo. 
Spiral arms are another cause for local streaming, which affect the~mass determination by several percent, while they do not  influence the~mass determination of the~dark halo much.   
 
There are several reviews on RCs and~mass determination of galaxies [\cite{SofueRubin2001,Sofue2017,Salucci2019}]. 
In this review, we revisit recent RC studies and~determination of the~local DM density in our Milky Way. 
 In Section \ref{Section2}, we briefly review the~current status of the~RC determinations along with the~methods. \rev{In Sections \ref{Section3} and \ref{Section4} we propose a~new method to use the~surface mass density (SMD) directly calculated from a~unified RC to estimate the~local DM density, and~apply it to the~newest RC of the~Milky Way up to radius of $\sim 100$ kpc. } 
We~adopt the~galactic constants: $(\rzero,\vzero)$=(8.0 kpc, 238 \kms) [\cite{Honma+2012,Honma+2015}], 
where $\rzero$ is the~distance of the~Sun from the~galactic center (GC) and~$\vzero$ is the~circular velocity of the~local standard of rest (LSR) at the~Sun [\cite{Fich+1991}].

\section{Rotation Curve of the~Milky Way}\label{Section2}

\vspace{-8pt}
\subsection{Progress in the~Last Decades}

The galactic RC is dependent on the~galactic constants. Accordingly, the~uncertainty and~error in the~RC include uncertainties of the~constants. Currently recommended,  determined,  or measured values are summarized in Table \ref{tabGal}, where they appear to be converging to around $\sim 8.0-8.3$ and~\mbox{$\sim 240$ \kms}. In this paper, we adopt $R_0= 8.0$ kpc and~$V_0 =238$ \kms from the~recent measurements with VERA
(VLBI Experiments for Radio Astrometry) 
[\cite{Honma+2012,Honma+2015}].

\begin{table}
\begin{center}
\caption{Galactic constants ($R_0,V_0$). } 
\label{tabGal}
\begin{tabular}{lll}
\hline
\hline
\textbf{Authors (Year)} & \boldmath{$R_0$} \textbf{(kpc)} & \boldmath{$V_0$} \textbf{(\kms)} \\ 
\hline
IAU recommended (1982)& 8.2  & 220  \\

Review before 1993 (Reid 1993)
[\cite{Reid1993}] &$8.0 \pm 0.5$   &  \\

Olling and~Dehnen 2003
[\cite{Olling+2003}] & $7.1\pm 0.4$ & $184\pm 8$ \\

VLBI Sgr A$^*$ (Ghez et al. 2008)
[\cite{Ghez+2008}]&$8.4 \pm 0.4$  &  \\

ibid (Gillessen et al. 2009)
[\cite{Gillessen+2009}]& $8.33 \pm 0.35$  &  \\

Maser astrometry (Reid et al. 2009)
[\cite{Reid+2009}] & $8.4\pm 0.6$  & $254\pm 16$  \\

Cepheids (Matsunaga et al. 2009)
[\cite{Matsunaga+2009}] & $8.24 \pm 0.42$ \\

VERA (Honma et al. 2012, 2015)
[\cite{Honma+2012,Honma+2015}]. & $8.05\pm 0.45$ & $238\pm 14$ \\

Adopted in this paper & 8.0 & 238\\
\hline
\end{tabular}    
\end{center}
\end{table} 
 
The RC of the~galaxy has been obtained by various methods as described in the~next subsection, and~many authors presented their results based on different galactic constants  (Table \ref{tabrcmw}).

\begin{table}
\begin{center}
\caption{Rotation curves (RCs) of the~Milky Way galaxy.}  
\label{tabrcmw}
\begin{tabular}{lll}
\hline
\hline
\textbf{Authors (Year)} & \textbf{Radii (kpc)} & \textbf{Method} \\ 
\hline 
Burton and~Gordon (1978)[\cite{Burton+1978}]& 0--8 & HI tangent \\
Blitz et al. (1979)
[\cite{Blitz+1979}]& 8--18 & OB-CO assoc.  \\
Clemens (1985)[\cite{Clemens1985}]& 0 -18 & CO/compil. \\
Dehnen  and~ Binney (1998)[\cite{Dehnen+1998}]& 8--20& compil. + model \\
Genzel et al. (1994--), Ghez et al. (1998--)[\cite{Genzel+2010,Ghez+2008}]&0--0.0001 &GC IR spectr. \\    
Battinelli, et al. (2013)[\cite{Battinelli+2013}]&  9--24 & C stars \\
Bhattacharjee et al.(2014)[\cite{Bhattacharjee+2014}]&  0--200& Non-disk objects \\
Lopez-Corredoira (2014)[\cite{Lopez2014}]& 5--16 & Red-clump giants $\mu$ \\
Boby et al. (2012)[\cite{Bovy+2012b}]&4-14&NIR spectroscopy\\
Bobylev (2013); --- \& Bajkova (2015)[\cite{Bobylev2013,Bobylev+2015}] & 5--12 & Masers/OB stars \\
Reid et al. (2014)[\cite{Reid+2014}]&4-16&Masers SF regions, VLBI \\
Honma et al. (2012, 2015)[\cite{Honma+2012,Honma+2015}]& 3--20 & Masers,VLBI \\
Iocco et al. (2015, 2016); Pato \& Iocco (2017a,b)[\cite{Iocco+2015,Iocco+2016,Pato+2017a,Pato+2017b}]& 1--25 kpc& CO/HI/opt/maser/compil.\\
Huang et al. (2016)[\cite{Huang+2016}]& 4.5--100 & HI/opt/red giants\\
Kre{\l}owski et al (2018)[\cite{Krelowski+2018}]& 8--12 & GAIA\\
Lin and~Li (2019)[\cite{Lin+2019}] & 4--100  & compil.\\
Eilers et al (2019)[\cite{Eilers+2019}]& 5--25 & Wise, 2Mass, GAIA\\
Mr\'oz et al. (2019)[\cite{Mroz+2019}]& 4--20& Classical cepheids\\
Sofue et al. (2009); Sofue (2013, 2015, this work)[\cite{Sofue+2009,Sofue2013,Sofue2015}]&0.01--1000 & CO/HI/maser/opt/compil.\\
\hline
\end{tabular}  
\end{center}
\end{table}

In the~1970--1980s, the~inner RC was extensively measured using the~terminal-velocities of HI (neutral hydrogen) and~CO (carbon monoxide) gases 
[\cite{Burton+1978, Clemens1985, Fich+1989}]. 
In the~late 1980s to the 2000s, outer~rotation velocities were measured by combining optical distances of OB  
[\cite{Blitz+1979, Demers+2007}].
The~HI thickness method was also useful to measure rotation of the~entire disk
[\cite{Merrifield1992, Honma+1997}]. 
The innermost mass distributions inside the~GC have been obtained extensively since the 1990s using the motion of infrared stellar objects
[\cite{Genzel+2010,Ghez+2008, Lindqvist+1992, Gillessen+2009}]. 
 
Trigonometric determinations of both the~3D positions and~velocities have provided the~strongest tool to date for measurement of the~galactic rotation  
[\cite{Honma+2007,Honma+2012,Honma+2015,Sakai+2015,Nakanishi+2015}].  
A number of optical parallax measurements of stars such with GAIA have been obtained for RC determination
[\cite{Lopez2014,Krelowski+2018}]. 

The total mass of the~galaxy, including the~extended dark halo, has been measured by analyzing the~outermost RC and~motions of satellite galaxies orbiting the~galaxy, and~the mass up to \mbox{$\sim 100-200$} kpc has been estimated to be $\sim 3 \times 10^{11}\Msun$
[\cite{Sofue2015,Callingham+2019}].

\subsection{Methods to Determine the~Galactic RC}

The particular location of the~Sun inside the~Milky Way makes it difficult to measure the~rotation velocity of the~galactic objects. Sophisticated methods have been developed to solve this problem, as~briefly described below.
  
\subsubsection{Tangent-Velocity Method } 
Inside the~solar circle ($-90^\circ \le l \le 90^\circ$), the~galactic gas disk has tangential points, at which the~rotation velocity is parallel to the~line of sight and~attains the~maximum radial velocity $ {\vr}_{\rm ~max}$ (terminal or  tangent-point velocity). The~rotation velocity $V(R)$ at galacto-centric distance $R=\Rsun \sin~ l$ is calculated simply correcting for the~solar motion. 

 \subsubsection{ Radial-Velocity + Distance Method } 
If the~distance $r$ of the~object is measured by spectroscopic and/or trigonometric observations, the~rotation velocity is obtained by geometric conversion of the~radial velocity, distance, and~the longitude. 
The distance has to be measured independently, often using spectroscopic distances of OB stars, and~the distances are assumed to be the~same as those of associated molecular clouds and~HII (ionized hydrogen) regions, whose radial velocities are observed by radio lines. Since the~photometric distances have often large errors, obtained RC plots show large scatter. 
 
 \subsubsection{Trigonometric Method} 
If the~proper motion and~radial velocity along with the~distance are measured at the~same time, or from different observations, the~3D velocity vector, and~therefore the~rotation velocity, of any source is uniquely determined without being biased by assumption of circular motion as well as the~galactic constants. VLBI  (very long baseline interferometer) measurements of maser sources
[\cite{Honma+2007,Honma+2012,Honma+2015,Nakanishi+2015}] 
and~optical/IR trigonometry of stars
[\cite{Roeser+2010,Lopez2014}] 
have given the~most accurate RC.  

\subsubsection{Disk-Thickness Method}  
The errors in the~above methods are mainly caused by the~uncertainty of the~distance measurements. This disadvantage is eased by the~HI-disk thickness method
[\cite{Merrifield1992,Honma+1997}]. 
The~angular thickness of the~HI disk along an~annulus ring is related to can be used to determine the~rotation velocity by combining with radial velocity distribution along the~longitude.  

\subsubsection{Pseudo-RC from Non-Disk Objects}
Beyond or outside the~galactic disk, globular clusters and~satellite galaxies are used to estimate the~pseudo-circular velocity from their radial velocities based on the~Virial theorem, assuming that their motions are at random, or the~ rotation velocity is calculated by $\Vrot \sim \sqrt{2} v_g$, where  $v_g$ is the~galacto-centric radial velocity. 
{On the~other hand, Huang et al. (2016)
[\cite{Huang+2016}] have recently employed more sophisticated, probably more reliable, method to solve the~Jeans equations for the~non-disk stars and~clusters.}
 
\subsection{Unified RC}

A RC covering a~wide region of the~galaxy has been obtained by compiling the~existing data by re-scaling the~distances and~velocities to the~common galactic constants \mbox{$(R_0, V_0)$=(8.0 kpc, 200 \kms)} [\cite{Sofue+2009}], 
and~later to (8.0 kpc, 238 \kms)
[\cite{Sofue2013,Sofue2017}]. 
 In these works, the~central RC inside the~GC has been obtained from analyses of the~kinematics of the~molecular gas and~infrared stellar motions as well as the~supposed Keplerian motion representing the~central massive black hole. Outer RC beyond $R\sim 30$ kpc has been determined from the~radial motions of satellite galaxies and~globular clusters.

The RC determination has been improved recently by compiling a~large amount of data from a~variety of spectroscopic as well as trigonometric measurements from radio to optical wavelengths. An extensive compilation of the~data of rotation velocities of the~galactic disk has been published recently, and~is available as an~internet data base
[\cite{Iocco+2015,Iocco+2016,Pato+2015a,Pato+2015b,Pato+2017a,Pato+2017b}]. 

\begin{figure}
\begin{center}   
\includegraphics[width=8.5cm]{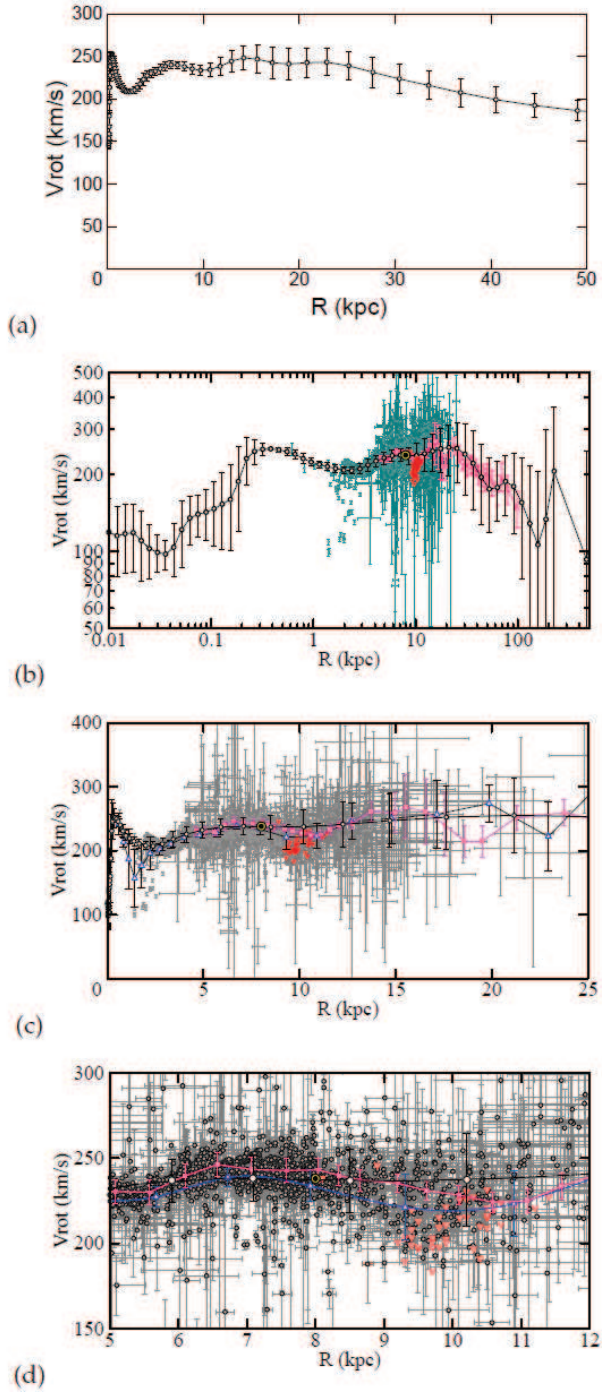}
\end{center} 
\caption{
 (\textbf{a}) Unified RC of the~Milky Way used in this paper for the~mass distribution obtained by averaging the~RCs from 
references [\cite{Sofue2015,Sofue2017,Huang+2016}]. 
The~bars are standard deviations within each Gaussian-averaging bin.
  The plotted values are listed in the tables in Appendix \ref{AppendixA}.
(\textbf{b}) Logarithmic RC of the~Milky Way from 
[\cite{Sofue2015,Sofue2017}] 
(circles), compared with those from the~recent literature: Green circles with error bars are from the~compilation by 
[\cite{Pato+2017a,Pato+2017b}]
and~blue triangles are their running averages. Red~triangles stand for data from [\cite{Krelowski+2018}]
based on GAIA data. 
{These two data are re-scaled to ($R_0, V_0$)=(8.0 kpc, 238 \kms).} 
Pink rectangles are the~RC by [\cite{Huang+2016}].
{without re-scaling}. (\textbf{c})~Same, but in linear scale.
(\textbf{d}) Same, but close up in the~solar vicinity.}  
\label{mwrc}
\end{figure}

Figure \ref{mwrc}a shows the~presently obtained unified RC using the~curves from[\cite{Sofue2015,Sofue2017}]
and~RC by Huang et al. (2016)[\cite{Huang+2016}] between $R=4.6$ and~$\sim 100$ kpc.
Although Huang et al. employed the~galactic constants of (8.34 pc, 240 \kms), we did not apply rescaling to (8.0, 238), because the~galacto-centric distances of off-plane objects are less dependent on the~solar position compared to the~disk objects as used for our RC at $<\sim 20$ kpc where the~rotation velocity is rather flat, and~also because their $V_0=240$ \kms is close to our 238 \kms. 
 
The unified RC was obtained by taking Gaussian running averages of rotation velocities from the~used RCs in each of newly settled radius bins, where the~ statistical weight of each input point was given by the~inverse of the~squared  error.

In Figure \ref{mwrc}b,c we compare the~unified RC with the~recent measurements by [\cite{Pato+2017a,Pato+2017b,Krelowski+2018}]
re-scaled to the~galactic constants of (8.0 kpc, 238 \kms) following the~method described in [\cite{Sofue+2009}].
Although individual data points are largely scattered, their averages well coincide with the~unified RC.
In the~figures we also compare the~data with the~RC by 
[\cite{Huang+2016}]
up to $\sim 100$ kpc without rescaling, which also coincides with the~other data within the~scatter.

 We here comment on the~property of the~unified RC built by averaging the~published data. It must be remembered that the~averaging procedure does not satisfy the~condition of statistics in the~strict meaning, because the~data are compiled from different authors using a~variety of instruments and~analysis methods, which makes it difficult to evaluate common statistical weights for the~used data points. So, remembering such a~property, in view that the~unified RC well approximates the~original curves as well as for its convenience for the~determination of the~mass distribution by the~least-squares and/or $\chi^2$ fitting, we shall employ it in our present analysis. 
  
\subsection{Mass Components}  

The rotation velocity is related to the~gravitational potential, hence to the~mass distribution, as
\be
V(R)=\sqrt{\Sigma V_i^2}=\sqrt{R  \frac{\partial \Sigma  \Phi_i }{ \partial R}},
\ee
where $\Phi_i$ is the~gravitational potential of the~$i$-th component and~$V_i$ is the~corresponding circular velocity. 
The rotation velocity is often represented by superposition of the~central black hole (BH), bulge, disk, and~the dark halo as
\be
V(R)= \sqrt{V_{\rm BH}(R)^2+V_{\rm b}(R)^2+V_{\rm d}(R)^2+V_{\rm h}(R)^2}.
\ee 

Here, the~subscript BH represents black hole, b stands for bulge, d for disk, and~h for the~dark halo. The~contribution from the~black hole can be neglected in sufficiently high accuracy, when the~dark halo is concerned.
The mass components are usually assumed to have the~following functional forms.
 
\subsubsection{Massive Black Hole}   
The GC of the~Milky Way is known to nest a~massive black hole of mass of $M_{\rm BH}\sim 4 \times 10^6\Msun$ 
[\cite{Genzel+2010,Ghez+2008,Gillessen+2009}].
The~RC is assumed to be expressed by a~curve following the~Newtonian potential of a~point mass at the~nucleus.
    
\subsubsection{De Vaucouleurs Bulge}   
The commonly used SMD profile to represent the~central bulge, which is assumed to be proportional to the~empirical optical profile of the~surface brightness, is the~\dv~law [\cite{deV1958}],
\be 
\Sigma_{\rm b}(R) 
= \Sigma_{\rm be} {\rm exp} \left[-7.6695 \left(\left( \frac{R}{ R_{\rm b}} \right)^{1/4}-1\right)\right],
\label{eq-smdb}
\ee  
where $ \Sigma_{\rm be} $ is the~value at radius $R_{\rm b}$ enclosing a~half of the~integrated surface mass [\cite{Sofue2017}].
Note that the~\dv surface profile, also the~exponential disk, has a~finite value at the~center. 
The volume mass density $\rho(r)$ at radius $r$ for a~spherical bulge is calculated using the~SMD by 
\be
\rho(r) = \frac{1}{\pi} \int_r^{\infty} \frac{d \Sigma_b(x)}{dx} \frac{1}{\sqrt{x^2-r^2}}dx,
\label{eq-rhob}
\ee 
 and~the mass inside $R$ is
\be
 M(R) =4\pi \int_0^R r^2\rho(r)dr.
 \ee  

The circular velocity is thus obtained by 
\be
 V_{\rm b}(R) = \sqrt{ \frac{GM(R)}{R}}.
 \ee 

More general form  
 $e^{-(R/r_e)^n}$ called the \mbox{the~S\'ersic} law 
 is discussed in relation to its dynamical relation to the~galactic structure based on the~more general profile [\cite{Ciotti1991,Trujillo2002}].
  
\subsubsection{Exponential Disk}
The galactic disk is generally represented by an~exponential disk [\cite{Freeman1970}],
where the~SMD is expressed as
\be 
\Sigma_{\rm d} (R)=\Sigma_{\rm d} {\rm exp}(-R/R_{\rm d}).
\label{eq-smdd}
\ee
 
Here, $\Sigma_{\rm d}$ is the~central value, $R_d$ is the~scale radius. The~total mass of the~exponential disk is given by $M_{\rm disk}= 2 \pi \Sigma_{dc} R_{\rm d}^2$.
The RC for a~thin exponential disk is expressed by [\cite{Binney+1987}]
\be
V_{\rm d}(R)
=\sqrt{4 \pi G \Sigma_0 R_{\rm d}y^2[I_0(y)K_0(y)-I_1(y)K_1(y)]},
\ee
where $y=R/ (2R_{\rm d}) $, and~$I_i$ and~$K_i$ are the~modified Bessel functions.

The dark halo is described in the~next section 

\section{Dark Halo}\label{Section3}

The existence of dark halos in spiral galaxies has been firmly evidenced from the~well established difference between the~galaxy mass predicted by the~luminosity and~the mass predicted by the~rotation velocities [\cite{SofueRubin2001,Sofue2017,Salucci2019}]. 

In the~Milky Way, extensive analyses of RC and~motions of non-disk objects such as globular clusters and~dwarf galaxies in the~Local Group have shown flat rotation up to $\sim 30$ kpc, beyond which the~RC declines smoothly up to $\sim 300$ kpc [\cite{Sofue2013,Sofue2015}].
Further analyses of non-disk tracer objects have also shown that the~outer RC declines in a~similar manner
[\cite{Bhattacharjee+2014,Huang+2016,Li+2017}].
The fact that the~rotation velocity beyond $R\sim 30$ kpc declines monotonically indicates that the~isothermal model can be ruled out in representing the~Milky Way's~halo.
 
 \subsection{Dark Halo Models}
 
There have been various proposed  DH models, which may be categorized into two types: The~cored halo models 
[\cite{Burkert1995,Salucci+2000,Brownstein+2006}]
are a modification of the~isothermal model with a~steeper decrease of density at large radii. The~central cusp models 
[\cite{Navarro+1995,Navarro+1997,Moore+1999,Fukushige+2004}]
are based on extensive $N$-body numerical simulations of the~structural evolution in the~cold dark matter scenario in the~expanding universe, which predict an~infinitely increasing central peak. In either type, all the~DH models predict decreasing DM density beyond $h$ as $\rho \propto R^{-3}$, or declining rotation velocity as $\Vrot \propto \sqrt{{\rm ln} \ R/R}$.

The cored halo models exhibit a central plateau of finite density with scale radius, or the core radius, $h$, and are often represented by the following functions, where $x=R/h$.

\noindent{\bf Isothermal halo}: 
\be 
\rho_{\rm Iso} (x)=\frac{\rho_{\rm Iso} ^0}{1+x^2},
\label{eq-iso} 
\ee  
\noindent{\bf  Beta model with $\beta=1$ } [\cite{Navarro+1995}] :
\be
\rho_\beta (x)=\frac{\rho_\beta ^0}{(1+x^2)^{3/2}} .
\ee
\noindent{\bf  Burkert model} [\cite{Burkert1995,Salucci+2000}] :
\be 
\rho_{\rm Bur} (x)=\frac{\rho_{\rm Bur} ^0}{(1+x)(1+x^2)},
\label{eq-bur}
\ee 
\noindent{\bf Brownstein model} [\cite{Brownstein+2006}] :
\be 
\rho_{\rm Bro} (x)=\frac{\rho_{\rm Bro} ^0}{1+x^3}.
\label{eq-bro}
\ee

On the other hand, the central cusp models are often represented by the following functions. 
\noindent{\bf NFW model} [\cite{Navarro+1996,Navarro+1997}] :
\be 
\rho_{\rm NFW} (x)=\frac{\rho_{\rm NFW}^0}{x(1+x)^2},
\label{eq-nfw}
\ee 
\noindent{\bf  Moore model} [\cite{Moore+1999,Fukushige+2004}]
~with $\alpha=1.5$:
\be 
\rho_{\rm Moo} (x)
=\frac{\rho_{\rm Moo}^0}{x^{\alpha}(1+x^{3-\alpha})}
=\frac{\rho_{\rm Moo}^0}{x^{1.5}(1+x^{1.5})}.
\label{eq-nfw}
\ee

\begin{figure} 
\begin{center} 
\includegraphics[width=10cm]{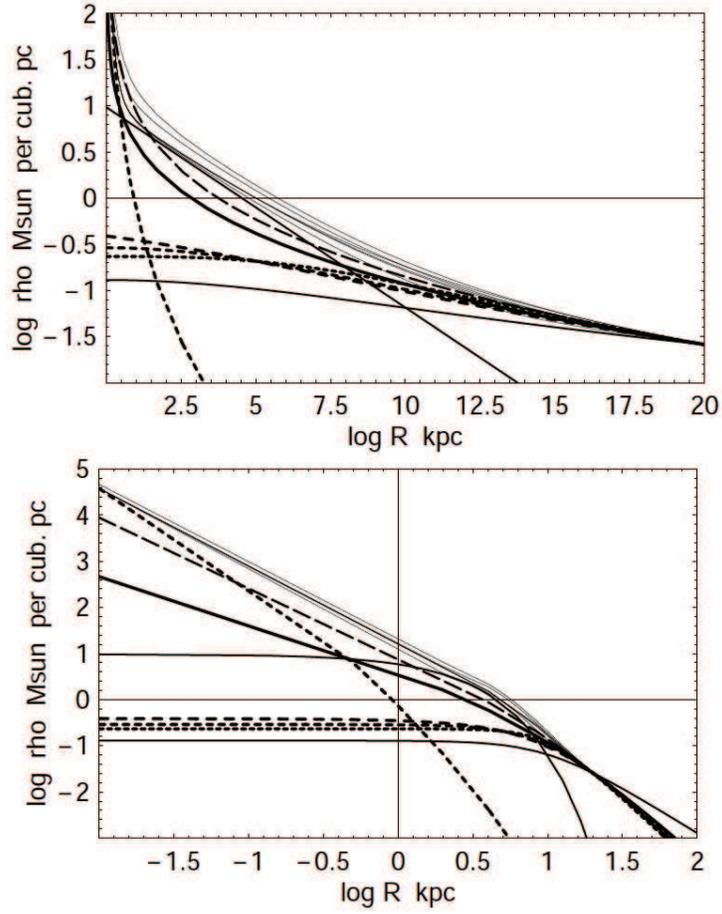} 
\end{center}
\caption{ (\textbf{Top}) Schematic density profiles of {NFW} (Navarro, Frenk, White) (thick solid), 
Moore (upper long dash), Burkert (long dash), Brownstein (dot), $\beta$  (dash), and~isothermal (thin solid) models with $h=10$ kpc normalized at 20 kpc, compared with the~disk (straight line) and~bulge (inner thick dash). Uppermost thin lines are the~sum of bulge, disk and~halo.  
(\textbf{Bottom}) Same, but in log--log plot. The~NFW cusp and~cored halos do not much contribute to the~mass density in the~GC, whereas the~Moore cusp somehow resembles the~bulge profile. }
\label{rhoModels} 
\end{figure}

 Figure \ref{rhoModels} shows schematic density profiles for various DH models with $h=10$ kpc combined with the~\dv bulge and~exponential disk, where the~halo density is normalized at $R=20$ kpc..

\subsection{Cusp vs Cored Halo}

The density profiles for the~NFW (Navarro, Frenk and White), Moore, Burkert, $\beta$, and~Brownstein models are almost identical beyond the~core radius $h$,  where they tend to $\propto R^{-3}$. Differences among the~models appear within the~Solar circle. 
The cusp models (NFW and~Moore models) predict steep increase of density toward the~center with a~singularity. The~cored halo models predict a~mild and~low density plateau in the~center with the~peak densities not much differing from each other within a~factor of two. However,~the~Burkert model has a~singularity with the~density gradient being not continuous across the~nucleus. 
 
Most of the~DH models predict lower density in the~innermost galaxy by two to several orders of magnitudes than the~bulge's density. This implies that the~DH does not much influence the~kinematics in the~inner galaxy. Namely, it is practically impossible to detect the~DM cusp by analyzing the~RC. Only the~Moore model predicts cusp density exceeding the~bulge's density in the~very center at $R<\sim 0.1$ pc, whereas the~applicability of the~model to such small sized region is not obvious [\cite{Fukushige+2004}].

\subsection{Central DM Density}

If we assume that the~functional form of the~NFW model is valid in the~very central region, the~SMD at $R\sim 100$ pc could be estimated to be about $\Sigma \sim 2.2\times 10^3\Msun {\rm pc}^{-2}$. 
This yields an~approximate volume density on the~order of $\rho \sim \Sigma/R\sim 11\Msun {\rm pc}^{-3} \sim 840$ GeV cm$^{-3}$ for a~detector of $\sim 1.4\deg$ resolution.

Such estimations could be a~key to the~indirect detection experiments of DM in the~GC \mbox{[\cite{Silk+1987,Escudero+2017}]
and~the literature therein)}. However, it is stressed that the~DM density in the~GC is two to several orders of magnitudes smaller than the~bulge's density on the~order of $10^4-10^5$ GeV cm$^{-3}$, making the~kinematical detection of DM difficult.

Interestingly, the~column density of DM, hence brightness (flux/steradian) of self-annihilation emission ($\gamma$-ray) stays almost constant against the radius and~is therefore constant regardless the~resolution of the~detector. On the~other hand, the~emission measure $\sim \rho^2 R$ varies as $\propto R^{-1}$, hence, the~brightness of collision-origin emission ($\gamma$ or microwave haze) increases toward the~center [e.g.,\cite{Finkbeiner2004}],
 so that the~detection rate will increase with the~detector's resolution.

Another concern about the~DM cusp is the~kinetic energy of individual particles. In order for the~cusp to be stationary, the~particles must be bound to the~gravitational potential, so that the~particle's speed must be lower than the~escaping velocity $v\sim \sqrt{2}\Vrot \sim 300$ \kms. 
This will give a~constraint on the~cross section $\sigma_A$ of the~DM annihilation, if the~collision rate $\sigma_A v$ is fixed by the~detection of DM-origin emissions.

The cored halo models (isothermal, Burkert, Brownstein, and~the $\beta$ models) predict a~mild and~finite-density plateau with scale radius of $h$ ($\sim 10$ kpc). Their central densities are also several orders of magnitude less than the~bulge's density, hence do not contribute to the~kinematics of the~gas and~stars in the~GC.
 
\rev{Finally, it is emphasized that the~discussion of a~central DM cusp has been obtained based on the~theoretical predictions, but neither on any observed phenomenon nor on measured quantity from the~RC or kinematics of GC objects. This makes a~strong contrast to the~observed fact of the~dark halo in the~outer galaxy, firmly evidenced by the~analyses of the~RC. 
}

\section{DM Density from Direct SMD}\label{Section4}

\vspace{-8pt}

\subsection{SMD from RC}

 In the~decomposition method of the~RC, the~resulting mass distribution depend on the~assumed functional forms of the~model profiles.
 In order to avoid this inconvenience, the~RC can be used to directly calculate the~surface mass distribution without employing any functional form. Only~an~assumption has to be made, either if the~galaxy's shape is a~sphere or a~flat disk.  
 
On the~assumption of spherical distribution, the~mass inside radius $R$ is given by
\begin{equation}
M(R)=\frac{R {V(R)}^{2}}{G}.
\label{masssphere}
\end{equation} 
 
Then the~surface-mass density (SMD) ${\Sigma}_{S}(R)$ at $R$ is calculated  by 
\be
\Sigma_{\rm S}(R) = 2 \int\limits_0^{\infty} \rho (r) dz,
\label{smdsphere}
\ee 
where  
\begin{equation}
\rho(r) =\frac{1}{4 \pi r^2} \frac{dM(r)}{dr}.
\label{rhosphere}
\end{equation}  

If the~galaxy is assumed to be a~flat thin disk, the~SMD ${\Sigma}_{\rm d}(R)$ is calculated by solving~Poisson's equation  (Freeman 1970; Binney  and~ Tremaine 1987) by
\begin{equation}
{\Sigma}_{\rm d}(R) =\frac{1}{{\pi}^2 G}  
\left[ \frac{1}{R} \int\limits_0^R 
{\left(\frac{dV^2}{dr} \right)}_x K \left(\frac{x}{R}\right)dx 
+ \int\limits_R^{\infty} {\left(\frac{dV^2}{dr} \right)}_x K \left
(\frac{R}{x}\right) \frac{dx}{x} \right].
\label{smdflat}
\end{equation}

Here, $K$ is the~complete elliptic integral, which becomes very large when $x\simeq R$. 

\begin{figure} 
\bc       
\includegraphics[width=10cm]{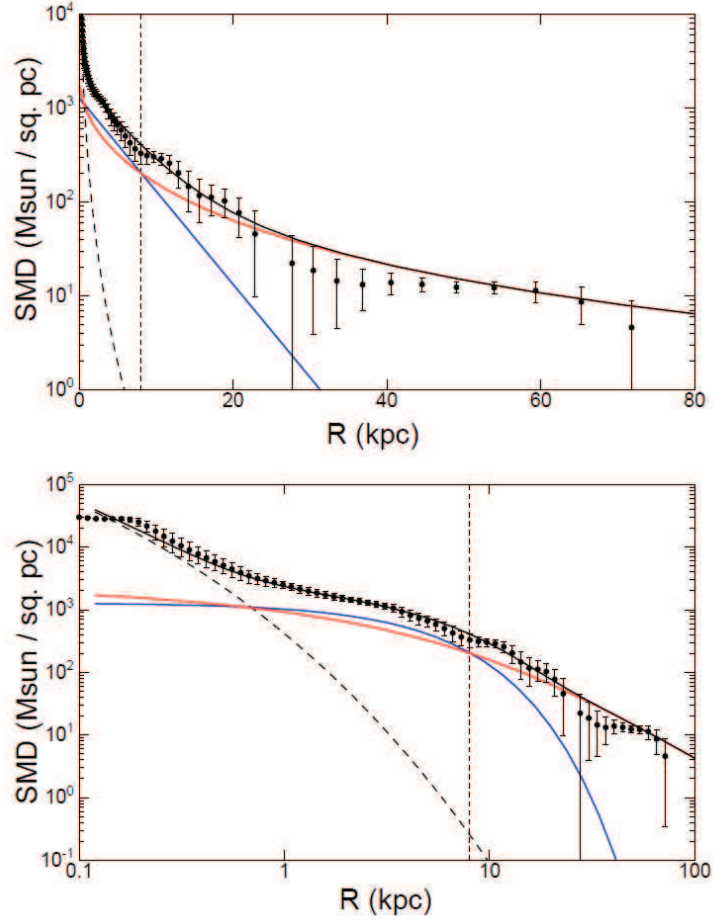}   
\ec
\caption{(\textbf{Top}) Direct surface-mass density (SMD) calculated for the~unified RC in figure \ref{mwrc} in spherical symmetry assumption (dots with error bars) in semi-logarithmic representation. The~solid line is the~$\chi^2$ fit, and~red, blue, and~dashed lines represent the~NFW halo, disk, and~bulge, respectively. (\textbf{Bottom}) Same, but in log--log plots. The~semi-logarithmic plot makes it easier to discriminate the~dark halo from exponential disk, which appears as a~straight line. The plotted values are listed in the tables in Appendix \ref{AppendixA}. }
\label{smd_fit} 
\end{figure}

The SMD distributions in the~galaxy for the~sphere and~flat-disk cases have been calculated for the~recent RCs [\cite{Sofue2017}].
In this paper we apply the~same method to the~here obtained unified RC (Figure \ref{mwrc}). Since we aim at studying the~dark halo, which is postulated to be rather spherical than a~flat disk, we assume spherical mass distribution. The~calculated SMD distribution is shown in Figure~\ref{smd_fit}.  

The SMD is strongly concentrated toward the~center, reaching a~value as high as $\sim 10^5 \Msun~{\rm pc}^{-2}$ within $R\sim 10$ pc, representing the~core of the~central bulge with the~extent of several hundred pc. It is followed by a~straightly declining profile from $R\sim 2$ to 8 kpc in the~semi-logarithmic plot, representing the~exponential nature of the~galactic disk. In the~outer galaxy beyond $\sim 8$ kpc, the~SMD profile tends to be displaced from the~straight disk profile, and~is followed by an~extended outskirt with a~slowly declining profile, representing a~massive halo extending to the~end of the~RC measurement at \mbox{$\sim 100$ kpc}. 
 
\subsection{Fitting by Bulge, Disk, and~Dark Halo} 

In order to separate the~dark halo from the~disk and~bulge components, the~well established RC decomposition method has been extensively applied to the~RCs 
[\cite{Sofue2017,Salucci2019}].
Besides this traditional method, we here propose to use the~SMD distribution. For this, we assume three mass components of \dv bulge, exponential disk, and~dark halo. In order to represent the, we employ the~NFW profile as a~'tool' for its popularity and~for the~dynamics background based on the~extensive numerical simulations.

We employ the~least $\chi^2$ fitting method, where $\chi^2$ is defined by  
\be
\chi^2=\Sigma_i[(SMD_i^{\rm direct} - SMD_i^{\rm calc})/\sigma_i]^2,
\ee
with $i$ denoting the~value at the~$i$-th data point, and~$\sigma_i$ is the~standard deviation around each data point in the~running averaging procedure of the~SMD distribution.

Fitting parameters are the~scale radius $a_d$ and~central SMD $\Sigma_d^0$ for the~disk, and~the scale (core) radius $h$ and~representative DM density $\rho_{\rm model}^0$ for the~halo.
 The bulge SMD is fixed to an~assumed \dv profile, which is negligible in the~present fitting range at $R\ge 1$ kpc. 

The fitting was obtained between $R=1$ and~100 kpc. The~fitting result for the~NFW halo model is shown in Figure \ref{smd_fit}. The~solid line is the~$\chi^2$ fit to SMD, and~red, blue, and~dashed lines represent the~halo, disk, and~bulge components, respectively. Note that the~semi-logarithmic plot makes it visually easier to recognize the~dark halo significantly displaced from the~exponential disk, which~appears as a~straight~line.

\rev{The present method to fit the~SMD distribution is essentially the same as that to fit the~RC in the~sense that both the~methods search for a~set of four parameters ($a_d$, $\Sigma_d$, $h$, and~$\rho_{\rm model}^0$) that minimize $\chi^2$ either of SMD or of RC.
However, an~advantage in using SMD is to make it visually easier  to discriminate the~disk from the~halo in the~semi-logarithmic plots (figure \ref{smd_fit}), where the~disk appears as a~straight~line. } 
 
\subsection{Local DM Density}

We thus obtained the~NFW DM halo parameters to be $h=10.94\pm 1.05$ kpc, $\rho_{\rm NFW}^0=0.787\pm 0.037$ \gevcc, which yields the local DM density $\rho_\odot=0.359\pm 0.017$ \gevcc. The~best-fit parameters for the~disk are determined to be $a_d=4.38\pm 0.35 $ kpc and~$\Sigma_0=(1.28\pm 0.09)\times 10^3 \Msun {\rm pc}^{-2}$. Table \ref{tab_fit} lists the~fitted result along with the~minimized $\chi^2$ value.  

\begin{table}
\begin{center}
\caption{Best-fit parameters of the~direct SMD by NFW halo and~exponential disk.}  
\label{tab_fit}  
\begin{tabular}{llll}  
\hline
\hline   
\textbf{Component} & \textbf{Parameter} & \textbf{Fitted Value} & \boldmath{$\chi^2$}\\ 
\hline
Expo. disk &$a_d$ & $4.38\pm 0.35 $ kpc \\
& $\Sigma_0$ & $(1.28\pm 0.09)\times 10^3 \Msun {\rm pc}^{-2}$\\
\hline 
NFW dark halo & $h$ & $10.94\pm 1.05$ kpc \\
& $\rho_{\rm NFW}^0$ & $0.787\pm 0.037$ \gevcc\\
 & $\rho_\odot$ & $0.359\pm 0.017$ \gevcc & 11.9 \\
 \hline 
Burkert$^\dagger$ & $\rho_\odot$ & $\sim 0.30\pm 0.02$ \gevcc & $17.3$\\
Brownstein$^\dagger$ & $\rho_\odot$ & $\sim 0.40\pm 0.02$ \gevcc& $17.9$\\
$\beta$ model$^\dagger$ & $\rho_\odot$ & $\sim 0.31\pm 0.02$ \gevcc &$17.3$\\
\hline
\end{tabular} 

$^\dagger$ Rough fitting, not conclusive.
\end{center}  
\end{table} 

We also obtained $\chi^2$ fitting using the~Burkert, Brownstein, and~$\beta$ profiles, and~listed the~local DM density and~minimized $\chi^2$  in Table \ref{tab_fit}. 
In these three models, the~$\chi^2 \sim 17-18$  were found to be systematically greater than that for the~NFW model ($\chi^2=11.9$). The~reason for the~difference is due to the~systematic difference in the~functional behavior between NFW and~the other three models: NSF has a~cusp steeply increasing toward the~center with sharpening scale radius, which results in the~possibility of finer fitting to the~slightly curved SMD profile at $R<\sim 10$ kpc in the~semi-log plot. On the~contrary, the~other three models predict  almost negligible SMD there, so that halo parameters contribute less intensively to the~fitting in the~innermost region, or the~fitting must be done only by the~disk's two parameters there, resulting in worse fitting.

\begin{table}[H] 
\caption{Current determinations of the~local DM density and~the literature.}
\label{tab_localdm}    
\begin{center}
\begin{tabular}{lllll}  
\hline
\hline  
{\bf Reference}  & \boldmath{$\rho_\odot$}  & \boldmath{$R_0$} & \boldmath{$V_0$} \\
& \boldmath{$({\rm GeV~cm}^{-3})^\dagger$} & \textbf{(kpc)} &  \textbf{(\kms)}\\
\hline
Weber and~de Boer (2010)[\cite{Weber+2010}] &   0.2 - 0.4 \\   
Catena and~Ulio (2010)[\cite{Catena+2010}]
 &$0.389 \pm 0.025$& && \\
Bovy and~Tremaine (2012) [\cite{Bovy+2012}]&   $0.3\pm 0.1$ \\
Piffl et al. (2014) [\cite{Piffl+2014}] &   0.58 \\ 
Pato et al (2015), Pato \& Iocco (2015) [\cite{Pato+2015a,Pato+2015b}] & $0.42\pm 0.25$  && 230 \\ 
Huang et al. (2016)[\cite{Huang+2016}] & $0.32 \pm 0.02$ &8.34 &240&\\
McMillan (2017)[\cite{McMillan2017}] & $0.38\pm 0.04$ &  8.21 & 233.1\\ 
Lin and~Li (2019)[\cite{Lin+2019}] & $0.51 \pm 0.09$ & 8.1 & 240 \\
Salucci et al. (2010, 2019) [\cite{Salucci+2010,Salucci2019}]&$0.43\pm 0.06$ &8.29&239\\
Eilers et al (2019) [\cite{Eilers+2019}]& $0.3\pm 0.03$ &  8.1 & 229\\
de Salas et al. (2019) [\cite{deSalas+2019}]&$0.3 - 0.4$ & \\
Cautun et al (2019) [\cite{Cautun+2019}] & $0.34\pm 0.02$&8 & 229 \\
Karukes et al (2019) [\cite{Karukes+2019}]& $0.43 \pm 0.02$ &8.34& 240\\
Sofue (2013)  [\cite{Sofue2013}]& $0.40 \pm 0.04$  &8.0 & 238 \\
----- (2020 this paper)&$0.36\pm 0.02$  &8.0 & 238 \\
\hline
Average$^\ddagger$ & $0.387 \pm 0.080$   \\
\hline
\end{tabular} 
 
$^\dagger$ \gevcc=38.2 $\Msun\ {\rm pc}^{-3}$. $^\ddagger$ Simple average of the~listed values with equal weighting.\\
\end{center}
\end{table}  

\begin{figure} 
\bc     
\includegraphics[width=9cm]{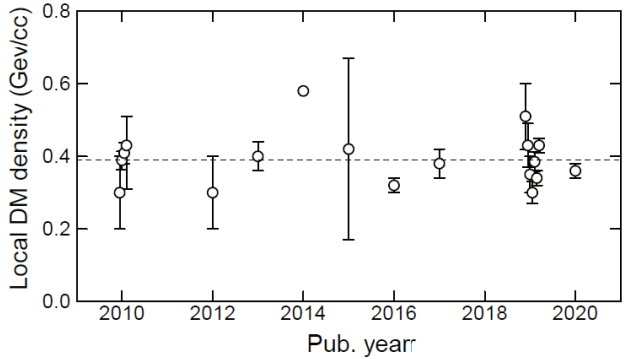}  
\ec
\caption{Local dark matter (DM) density from the~literature (Table \ref{tab_localdm}) plotted against publication year. The~dashed line indicates a~simple mean of the~plots at $\rho_\odot=0.39 \pm 0.09$ \gevcc.}
\label{localDMauthors}  
\end{figure}

 The local DM density is a~key quantity in laboratory experiments by the~direct detection of DM, and~has been estimated by a~number of authors with a~variety of methods. 
In Table \ref{tab_localdm} we list the~local DM densities from the~literature  along with the~present value for NFW profile. They are also plotted in Figure \ref{localDMauthors} against publication years. The~$\rho_\odot$ values seem to be nearly constant in the~decade. Averaging all the~listed values with an~equal weighting yields  $\rho_\odot=0.39 \pm 0.09$ \gevcc, which may be taken as a~'canonical' value.

\subsection{Dependence on the~Galactic Constants}

We have re-scaled the~adopted RC to $(R_0,V_0)=(8.0, 238)$ (kpc, \kms), which may vary within several \%. The~resulting local DM density will  vary accordingly, depending on the~constants. The~local mass density of the~spherical component is dependent on the~constants as $\rho_0 \propto R_0 V_0^2/R_0^3 \sim V_0^2 R_0^{-2}$. For small corrections $\delta V_0$ and~$\delta R_0$, the~DM density will change as $\delta \rho_0/\rho_0 \sim 2(\delta V_0/ V_0-\delta R_0/R_0)$. For~example, for $\delta V_0 \sim \pm $ 10 \kms, the~estimated local density varies by $\delta \rho_0/\rho_0 \sim \pm 0.08$, or for $\delta R_0 \sim \pm 0.1$ kpc, $\delta \rho_0/\rho_0 \sim \mp 0.025$.

                \section{Summary}
 
We reviewed the~current status of determination of the~RC of the~Milky Way, and~presented a~unified RC from the~GC to outer halo at $R\sim 100$ kpc. The~RC was used to directly calculate the~SMD without assuming any functional form. The~disk appears as a~straight line on the~semi-logarithmic plot of SMD against $R$, and~is visually well discriminated from the~DH having an~extended outskirt. 

The SMD distribution was fitted by a~bulge, disk, and~NFW dark halo using the~$\chi^2$ method. The~best-fit DH profile yielded the~local DM density of $0.359 \pm 0.017$ \gevcc. We also reviewed the~current estimations from the~literature in the~last decade, which appear to be converging to a~mean value of $\rho_\odot=0.39 \pm 0.09$ \gevcc. 

\vspace{6pt}

{\noindent {\bf Acknowledgments}} {The data analysis was performed at the~Center of Astronomical Data Analysis of the~National Astronomical Observatory of Japan.  
The author is grateful to Professor A. Hofmeister for inviting him to this special~issue.
} 
 
\appendix

\section{Tables concerning the RC and~SMD of the~Milky Way}
\label{AppendixA} 

Tables \ref{tabrcA} and~\ref{tabrcB} list the~running-averaged RC of the~Milky Way using the~data from [\cite{Sofue2015,Sofue2017,Huang+2016}],
which is used to calculate the~SMD in Figure \ref{smd_fit}. Tables \ref{tab_smdsA} and~\ref{tab_smdsB} lists the~directly calculated SMD from the~RC

\begin{table}[H]
\begin{center}
\caption{Rotation curve of the~Milky Way used in Figure \ref{mwrc}.} 
\label{tabrcA}
\scalebox{0.7}[0.7]{
\begin{tabular}{ccc}  
\hline 
\hline
\textbf{Radius} & \boldmath{$\Vrot$}  & \textbf{Standard Dev}.  \\ 
 (kpc) & (\kms) &  (\kms) \\ 
\hline

      0.100 &      144.9 &        3.7 \\
      0.110 &      147.4 &        4.2 \\
      0.121 &      150.4 &        4.8 \\
      0.133 &      153.8 &        6.1 \\
      0.146 &      158.9 &       10.3 \\
      0.161 &      167.4 &       16.1 \\
      0.177 &      180.1 &       22.4 \\
      0.195 &      196.6 &       27.1 \\
      0.214 &      213.6 &       26.9 \\
      0.236 &      227.8 &       22.7 \\
      0.259 &      237.9 &       17.0 \\
      0.285 &      244.4 &       11.8 \\
      0.314 &      248.2 &        7.6 \\
      0.345 &      250.2 &        4.7 \\
      0.380 &      251.0 &        2.9 \\
      0.418 &      250.7 &        2.1 \\
      0.459 &      249.7 &        2.3 \\
      0.505 &      248.0 &        2.9 \\
      0.556 &      245.9 &        3.7 \\
      0.612 &      243.2 &        4.6 \\
      0.673 &      239.8 &        5.7 \\
      0.740 &      235.8 &        6.4 \\
      0.814 &      231.7 &        6.5 \\
      0.895 &      227.8 &        6.0 \\
      0.985 &      224.5 &        5.2 \\
      1.083 &      221.7 &        4.5 \\
      1.192 &      219.1 &        4.0 \\
      1.311 &      216.8 &        3.7 \\
      1.442 &      214.7 &        3.4 \\
      1.586 &      212.7 &        3.1 \\
      1.745 &      210.9 &        2.8 \\
      1.919 &      209.5 &        2.3 \\
      2.111 &      208.5 &        1.8 \\
      2.323 &      208.2 &        1.6 \\ 
\hline 
\end{tabular} 
} 
\end{center}
\end{table}  

\begin{table}[H] 
\begin{center}
\caption{Continued from Table \ref{tabrcA}.} 
\label{tabrcB}
\scalebox{0.7}[0.7]{
\begin{tabular}{ccc}  
\hline
\hline
\textbf{Radius} & \boldmath{$\Vrot$}  & \textbf{Standard Dev}.  \\ 
 (kpc) & (\kms) &  (\kms) \\ 
\hline 
      2.555 &      208.9 &        2.2 \\
      2.810 &      210.7 &        3.6 \\
      3.091 &      213.4 &        4.8 \\
      3.400 &      217.2 &        5.9 \\
      3.740 &      222.0 &        6.6 \\
      4.114 &      226.6 &        5.7 \\
      4.526 &      229.5 &        4.4 \\
      4.979 &      231.6 &        4.3 \\
      5.476 &      234.1 &        5.3 \\
      6.024 &      237.2 &        5.7 \\
      6.626 &      239.5 &        5.0 \\
      7.289 &      240.1 &        4.1 \\
      8.018 &      239.0 &        4.4 \\
      8.820 &      236.7 &        5.4 \\
      9.702 &      234.5 &        6.0 \\
     10.672 &      234.2 &        7.1 \\
     11.739 &      237.1 &        9.8 \\
     12.913 &      242.8 &       12.4 \\
     14.204 &      248.5 &       13.3 \\
     15.625 &      249.7 &       14.8 \\
     17.187 &      246.2 &       17.4 \\
     18.906 &      243.3 &       18.3 \\
     20.797 &      243.9 &       17.5 \\
     22.876 &      245.6 &       15.6 \\
     25.164 &      243.7 &       15.2 \\
     27.680 &      237.3 &       16.1 \\
     30.448 &      229.6 &       15.5 \\
     33.493 &      222.5 &       14.1 \\
     36.842 &      215.0 &       14.0 \\
     40.527 &      207.1 &       13.8 \\
     44.579 &      200.3 &       12.7 \\
     49.037 &      194.7 &       11.9 \\
     53.941 &      189.8 &       11.3 \\
     59.335 &      186.2 &       10.4 \\
     65.268 &      184.7 &        9.6 \\
     71.795 &      183.9 &        9.3 \\
     78.975 &      181.4 &       11.0 \\
     86.872 &      175.5 &       14.6 \\
     95.560 &      167.7 &       16.3 \\
\hline
\end{tabular}  }
\end{center}
\end{table}  
                
\begin{table}[H] 
\begin{center}
\caption{Directly calculated SMD by spherical assumption of the~mass distribution.} 
\label{tab_smdsA}
\scalebox{0.7}[0.7]{
\begin{tabular}{ccc}  
\hline
\hline
\textbf{Radius}  & \boldmath{$SMD$} & \textbf{Standard Dev.} \\ 
\textbf{(kpc)} & \boldmath{$\Msun \ {\rm pc}^{-2}$} & \boldmath{$\Msun \ {\rm pc}^{-2}$} \\ 
\hline
      0.100 &      29933.0 &        861.3 \\
      0.110 &      29054.0 &        654.8 \\
      0.121 &      28384.0 &        666.0 \\
      0.133 &      28160.0 &        570.8 \\
      0.146 &      28319.0 &        637.2 \\
      0.161 &      28203.0 &       1406.6 \\
      0.177 &      27368.0 &       2481.1 \\
      0.195 &      25014.0 &       3514.2 \\
      0.214 &      21548.0 &       4357.7 \\
      0.236 &      17908.0 &       4806.8 \\
      0.259 &      14804.0 &       4733.1 \\
      0.285 &      12369.0 &       4231.5 \\
      0.314 &      10489.0 &       3549.3 \\
      0.345 &       8978.9 &       2929.7 \\
      0.380 &       7736.5 &       2384.1 \\
      0.418 &       6700.8 &       1959.1 \\
      0.459 &       5830.8 &       1636.9 \\
      0.505 &       5090.6 &       1374.2 \\
      0.556 &       4452.0 &       1158.1 \\
      0.612 &       3899.9 &        973.5 \\
      0.673 &       3464.9 &        803.4 \\
      0.740 &       3145.2 &        644.7 \\
      0.814 &       2904.3 &        510.8 \\
      0.895 &       2701.7 &        415.1 \\
      0.985 &       2510.1 &        354.4 \\
      1.083 &       2320.9 &        319.0 \\
      1.192 &       2144.8 &        291.9 \\
      1.311 &       1985.0 &        266.7 \\
      1.442 &       1843.6 &        240.7 \\
      1.586 &       1718.4 &        214.5 \\
      1.745 &       1611.4 &        188.2 \\
      1.919 &       1519.3 &        164.5 \\
      2.111 &       1440.6 &        144.4 \\
      2.323 &       1368.3 &        130.9 \\
 
\hline
\end{tabular}}
\end{center}
\end{table}   
               
\begin{table}[H] 
\begin{center}
\caption{Continued from Table \ref{tab_smdsA}.} 
\label{tab_smdsB}
\scalebox{0.7}[0.7]{
\begin{tabular}{ccc}  
\hline
\hline
\textbf{Radius}  & \boldmath{$SMD$} &Standard Dev. \\ 
(kpc) & $\Msun \ {\rm pc}^{-2}$ & $\Msun \ {\rm pc}^{-2}$ \\ 
\hline
     
      2.555 &       1296.4 &        125.1 \\
      2.810 &       1220.5 &        126.3 \\
      3.091 &       1139.8 &        134.1 \\
      3.400 &       1055.2 &        146.1 \\
      3.740 &        944.3 &        157.4 \\
      4.114 &        824.6 &        161.5 \\
      4.526 &        734.9 &        150.5 \\
      4.979 &        668.3 &        133.7 \\
      5.476 &        600.8 &        123.8 \\
      6.024 &        523.5 &        119.3 \\
      6.626 &        446.8 &        113.6 \\
      7.289 &        383.7 &        101.6 \\
      8.018 &        339.1 &         83.4 \\
      8.820 &        314.1 &         62.2 \\
      9.702 &        303.4 &         43.9 \\
     10.672 &        293.2 &         39.0 \\
     11.739 &        272.0 &         48.2 \\
     12.913 &        229.5 &         58.5 \\
     14.204 &        170.5 &         65.2 \\
     15.625 &        127.5 &         62.4 \\
     17.187 &        114.7 &         46.4 \\
     18.906 &        110.0 &         35.2 \\
     20.797 &         91.8 &         33.5 \\
     22.876 &         61.2 &         34.5 \\
     25.164 &         37.2 &         33.2 \\
     27.680 &         28.0 &         25.4 \\
     30.448 &         24.7 &         17.1 \\
     33.493 &         20.3 &         11.6 \\
     36.842 &         17.3 &          7.8 \\
     40.527 &         17.1 &          4.7 \\
     44.579 &         16.7 &          3.0 \\
     49.037 &         15.2 &          2.3 \\
     53.941 &         14.4 &          2.3 \\
     59.335 &         13.4 &          3.3 \\
     65.268 &         10.6 &          4.3 \\
     71.795 &          6.2 &          4.9 \\

\hline
\end{tabular} }
\end{center}
\end{table}   


\end{document}